\documentclass[aps,pra,showpacs,twocolumn]{revtex4}

\usepackage{amssymb}
\usepackage{epsfig}
\usepackage{graphicx}
\usepackage{amsmath}
\usepackage{array,color}
\usepackage{dcolumn}
\usepackage{bm}

\begin{document}

\title{An Artificial Frustrated System: Cold Atoms in 2D Triangular Optical Lattice}
\author{Yao-Hua Chen, Wei Wu, Hong-Shuai Tao, and Wu-Ming Liu}
\address{Beijing National Laboratory for Condensed Matter Physics,
Institute of Physics, Chinese Academy of Sciences, Beijing 100190,
China}

\date{\today}

\begin{abstract}
We investigate the strongly correlated effect of cold atoms in
triangular optical lattice by dynamical cluster approximation
combining with continuous time quantum Monte Carlo method. When
the interaction increases, Fermi surface evolves from a circular
ring into a flat elliptical ring, system translates from Fermi
liquid into Mott insulator. The transition between Fermi liquid
and pseudogap shows a reentrant behavior due to Kondo effect. We
give an experimental protocol to observe these phenomena by
varying lattice depth and atomic interaction via Feshbach
resonance in future experiments.
\end{abstract}


\maketitle

\section{Introduction}
Quantum phase transition in strongly correlated system is an
important research area in condensed matter physics, presenting
some of the most challenging problems. In a real material, the
experimental parameter is hard to be varied to observe the
strongly correlated effect, which is complicated by impurities and
multiple bands. However, a new developing technology called
optical lattices presents a highly controllable and clean system
for studying strongly correlated system, in which the relevant
parameter can be adjusted independently \cite{Jaksch, Petsas,
Immanuel, Hofstetter, Inada, Kottke}. Optical lattices with
different geometrical property can be set up by adjusting the
propagation directions of laser beams, such as triangular
\cite{Joksch}, honeycomb \cite{Duan} and kagom\'e \cite{Damch,
Santos, Damski2, Ruostekoski} optical lattice. The interaction
between the trapped atoms is tunable through the Feshbach
resonance, such as $^{6}Li$ and $^{40}K$. In recent years, a
series of experiments have been carried out to investigate the
quantum phase transition of cold atoms in optical lattice
\cite{Greiner, Hellweg, Jordans, Shin, Schneider, Gemelke, Greif}.

There are many analytical and numerical methods to investigate the
strongly correlated system, especially the frustrated system
\cite{TO, Takuya, Dimitrios, Takuma, Peng-Bin, Aryanpour,
Bhongale, Konotop, Yasuyuki, ZDLi}. The dynamical mean-field
theory (DMFT) \cite{Antoine} has been proved to be a useful tool.
The self-energy is given as a local quantity in DMFT, which has
been proved to be exact in the infinity dimensional limit
\cite{Metzner}. This method is a good approximation even in three
dimensional situation \cite{Helmes}. However, in the frustrated
systems, the nonlocal correlations can not be simply ignored, DMFT
would not work efficiently. So, many methods have been improved to
incorporate nonlocal correlations in the framework of DMFT, such
as dynamical cluster approximation (DCA) \cite{Thomas,Jarrell63}.
Different with the DMFT, the lattice problem is mapped into a
self-consistently embedded finite-sized cluster in DCA. The
irreducible quantities of the embedded cluster is used as an
approximation for the corresponding lattice quantities. DCA has
been used to investigate the geometrical frustrated system
\cite{Yoshiki, Lee}.

\begin{figure}[!t]
\centering
\includegraphics[width=8cm]{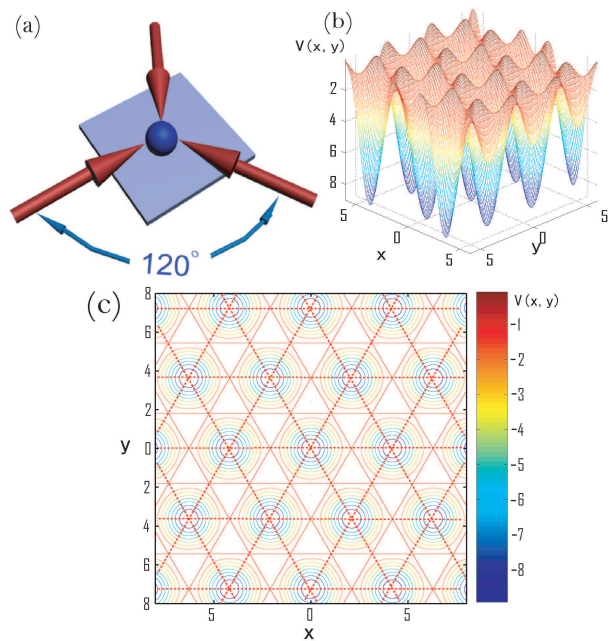}\hspace{0.8cm}
\caption{\label{fig:epsart}(Color online) (a) Sketch of
experimental setup to form triangular optical lattice. Each arrow
depicts a laser beam, the sphere in center of the figure depicts
the fermionic quantum gas, such as $^{40}K$. (b) Landscape of
potential $V(x,y)$. (c) The contour lines of triangular optical
lattice. The dark blue circles indicate the minimum lattice
potential. The dash red lines show the geometry of this triangular
optical lattice by connecting the minimum lattice potential. }
\end{figure}

In present Letter, we investigate the Mott transition in an
artificial frustrated system -- cold atoms in triangular optical
lattice. We improve the numerical method -- DCA combining with the
continuous time quantum Monte Carlo method (CTQMC) \cite{Rubtsov}
to investigate this geometrical frustrated system. CTQMC, used as
the impurity solver, is an exact numerical method proposed
recently. By calculating the density of states and the spectral
function, we find the system undergoes a second order phase
transition from Fermi liquid to Mott insulator. The phase diagram
shows a reentrant behavior of the transition between Fermi liquid
and pseudogap due to the Kondo effect at low temperature. These
phenomena can be observed in the triangular optical lattice by
varying the lattice depth and the interaction strength via the
Feshbach resonance. The advantage of this artificial frustrated
system is highly controllable and cleanness. And the numerical
method acts as a powerful tool for investigating the strongly
correlated effect in condensed matter physics, such as the spin
liquid, the high-temperature superconductor and the colossal
magnetoresistance.

\begin{figure}[t]
\centering
\includegraphics[width=6.0cm]{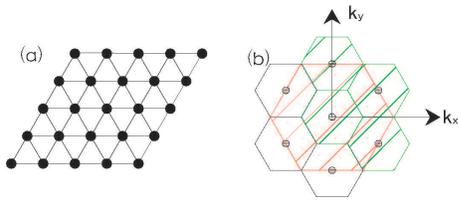}\hspace{0.8cm}
\caption{\label{fig:epsart}(Color online)(a) Sketch of triangular
lattice. (b) Coarse graining procedure in the first Brillouin Zone
when $N_c=4$. The red part shows the first Brillouin Zone in
reciprocal space. The green part shows the region divided into 4
cells for DCA calculation.}
\end{figure}

\section{Artificial Frustrated System}
In contrast
with real materials, the cold atoms trapped in an optical lattice
provide an artificial system to investigate the strongly
correlated effect. The experiment can be performed with $^{40}K$
atoms prepared by mixing two magnetic sublevels of the $F=9/2$
hyperfine manifold, such as the $|-9/2\rangle$ and the
$|-5/2\rangle$ states \cite{Jordans}. As an artificial frustrated
system, the triangular optical lattice can be set up by three
laser beams, such as the Yb fiber laser at wavelength
$\lambda=1064$ nm, with a $2\pi/3$ angle between each other, as
illustrated in Fig. 1(a). The potential of optical lattice is
given by
\begin{equation}\label{eq:eps}
V(x,y)=
V_0(3+4\cos(\frac{k_xx}{2})\cos(\frac{\sqrt{3}k_yy}{2})+2\cos(\sqrt{3}k_yy)),
\end{equation}
where $V_0$ is the barrier height of standing wave formed by laser
beams in the x-y plane, $k_x$ and $k_y$ are the two components of
wave vector $k=2\pi/\lambda$ along x and y directions. In
experiments, $V_0$ is always given in units of recoil energy
$E_r=\hbar^2k^2/2m$. The landscape of potential of triangular
optical lattice in the x-y plane is shown in Fig. 1(b), where the
dark blue parts in the figure indicate the minimum lattice
potential. Fig. 1(c) shows the contour lines of the triangular
optical lattice. By connecting the center of the circular which
indicates the minimum lattice potential, we may get the geometry
of this triangular optical lattice, as shown by the red dash
lines.

The Hamilitonian of the interacting fermionic atoms trapped in
this artificial frustrated system is written as
\begin{equation}\label{eq:eps}
H=-t\sum_{<ij>\sigma}c_{i\sigma}^{+}c_{j\sigma}+U\sum_{i}n_{i\uparrow}n_{i\downarrow},
\end{equation}
where $c_{i\sigma}^{+}$ and $c_{i\sigma}$ denote the creation and
the annihilation operator of the ferminic atom on lattice site i
respectively, $n_{i\sigma}=c_{i\sigma}^{+}c_{i\sigma}$ represents
the density operator of ferminic atom. And
$t=(4/{\sqrt{\pi}})E_r(V_0/E_r)^{3/4}exp(-2(V_0/E_r)^{1/2})$ is
the kinetic energy, which can be adjusted by the lattice depth
$V_0$. $U=\sqrt{8/\pi}ka_sE_r(V_0/E_r)^{3/4}$ is the on-site
interaction determined by the s-wave scattering length $a_s$,
which can be adjusted by Feshbach resonance \cite{Immanuel,
Duan05, Cheng}.

\begin{figure}[t]
 \centering
\includegraphics[width=6.0cm]{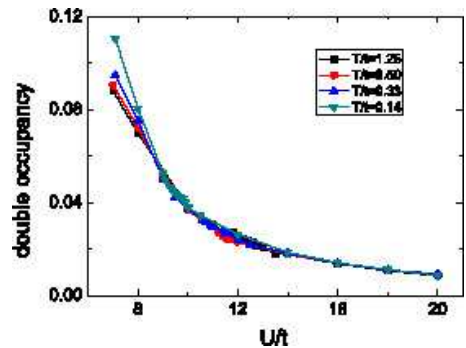}\hspace{0.5cm}
\caption{\label{fig:epsart}(Color online) The double occupancy
$D_{occ}$ as a function of the interaction U for different
temperature. $t$ is the kinetic energy in equation (2).}
\end{figure}

\section{Numerical Method: DCA+CTQMC}
We improve the dynamical cluster approximation (DCA) to combine
with the continuous time Quantum Monte Carlo method (CTQMC) to
investigate the strongly correlated effect of cold atoms in the
frustrated system shown in Fig. 2(a), which can be realized by
Hubbard model (2).

In DCA, the reciprocal space of the lattice containing $N$ points
is divided into finite cells \cite{Jarrell63}. The coarse-graining
Green's function $\overline{G}$ is achieved by averaging Green's
function $G$ within each cell. The lattice problem is mapped into
a self-consistently embedded finite-sized cluster. The coarse
graining procedure of the DCA is illustrated as follows: the
Brillouin zone is divided into $N_c$ cells, each cell is
represented by a cluster momentum $\overrightarrow{\mathbf{K}}$.
In Fig. 2(b), we provide an example of this coarse graining
procedure in $N_c=4$ situation. In our treatment, the
coarse-grained Green's function
\begin{eqnarray}
&&\overline{G}(\overrightarrow{\textbf{K}},i\omega_n)\nonumber=\frac{Nc}{N}\sum_{\widetilde{\textbf{k}}}G(\overrightarrow{\textbf{K}}
+\widetilde{\textbf{k}},i\omega_n)\nonumber\\
&&=\frac{Nc}{N}\sum_{\widetilde{\textbf{k}}}\frac{1}{i\omega_n-\varepsilon_{\overrightarrow{\textbf{K}}+
\widetilde{\textbf{k}}}-\overline{\Sigma_\sigma}(\overrightarrow{\textbf{K}},i\omega_n)},
\end{eqnarray}
 where summation over $\widetilde{\textbf{k}}$ is taken within
the coarse-graining cell, the $\omega_n$ is the Matsubara
frequency.

\begin{figure}[t]
\centering
\includegraphics[width=8cm]{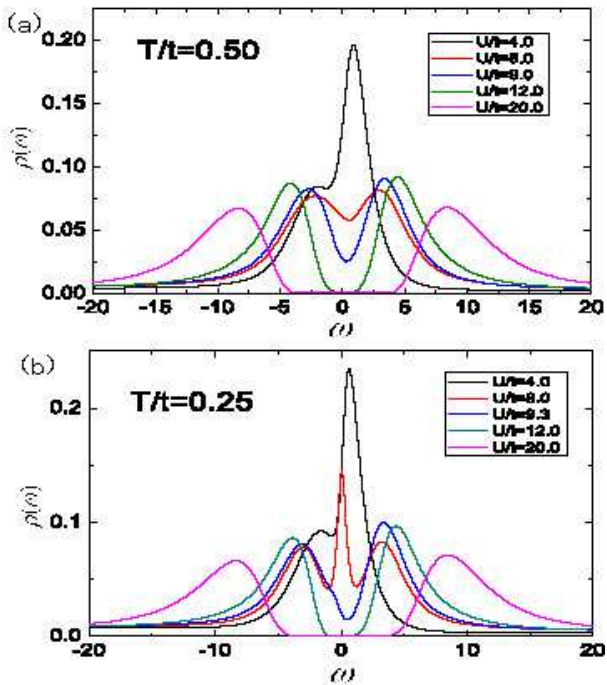}\hspace{0.5cm}
\caption{\label{fig:epsart}(Color online) The density of states
(DOS) as a function of frequency $\omega$ for different
interaction. $t$ is the kinetic energy in equation (2). (a) At
$T/t=0.50$, a pseudogap formed by the splitting of
Fermi-liquid-like peak appears when the interaction increases. A
gap is opened when the interaction is stronger than the critical
interaction $U_c/t=11.6$. (b) At $T/t=0.25$, a Kondo resonance
peak is found before the pseudogap appears.}
\end{figure}

Similar with DMFT, after mapping the Hubbard model to an Anderson
impurity problem, we introduce an impurity solver to solve the
cluster problem, such as quantum Monte Carlo (QMC), fluctuation
exchange approximate (FLEX), and the noncrossing approximation
(NCA). In our calculation, we employ the CTQMC \cite{Rubtsov}
which does not need to introduce any anxiliary-field variables as
our impurity solver. Comparing with the traditional QMC method,
the CTQMC is much more exact, because it does not use the Trotter
decomposition. We use  $10^7$ sweeps in our CTQMC step.

The self-consist loop can be taken as follows:

1) The DCA iteration loop can be started by setting the initial
self-energy $\Sigma_c(\overrightarrow{\textbf{K}},i\omega_n)$
which can be guessed or be get from a perturbation theory.

2) We could get:

$\overline{G}(\overrightarrow{\textbf{K}},i\omega_n)=\frac{Nc}{N}\sum_{\widetilde{\textbf{k}}}1/(i\omega_n-\varepsilon_{\overrightarrow{\textbf{K}}+
\widetilde{\textbf{k}}}-\overline{\Sigma_\sigma}(\overrightarrow{\textbf{K}},i\omega_n))$.

3) The host Green's function
$\overline{\mathcal{G}}(\overrightarrow{\textbf{K}},i\omega_n)$ is
computed by
$\overline{\mathcal{G}}(\overrightarrow{\textbf{K}},i\omega_n)^{-1}=\overline{G}(\overrightarrow{\textbf{K}},i\omega_n)^{-1}+\Sigma_c(\overrightarrow{\textbf{K}},i\omega_n).$

4) The
$\overline{\mathcal{G}}(\overrightarrow{\textbf{K}},i\omega_n)$ is
transformed from momentum-frequency variable to space-time
variable
$\overline{\mathcal{G}}(\overrightarrow{\textbf{X}}_i-\overrightarrow{\textbf{X}}_j,\tau_i-\tau_j)$
used as the input to the CTQMC simulation.

5) The CTQMC step is the most time consuming part of the iteration
loop. In our CTQMC step, we use $10^7$ CTQMC sweeps. After the
simulation, we get
$\overline{G}(\overrightarrow{\textbf{X}}_i-\overrightarrow{\textbf{X}}_j,\tau_i-\tau_j)$.

6)
$\overline{G}(\overrightarrow{\textbf{X}}_i-\overrightarrow{\textbf{X}}_j,\tau_i-\tau_j)$
is transformed from space-time variable to momentum-frequency
variable $\overline{G}(\overrightarrow{\textbf{K}},i\omega_n)$ by
Fourier transform.

7) We get the new self-energy by
$\Sigma_c(\overrightarrow{\textbf{K}},i\omega_n)=\overline{\mathcal{G}}(\overrightarrow{\textbf{K}},i\omega_n)^{-1}-\overline{G}(\overrightarrow{\textbf{K}},i\omega_n)^{-1}.$

8) Repeat from step 2) to step 7) until
$\Sigma_c(\overrightarrow{\textbf{K}},i\omega_n)$ converges to
desired accuracy.

9) Once convergence is reached, we can calculate the state density
$\rho(\omega)$ by maximum entropy method \cite{Gubernatis}. And we
can get other lattice quantities by some other additional analysis
code.

By combining the DCA, which introduces the nonlocal correlations,
and the exact numerical method: CTQMC, we can investigate the
frustrated system efficiently.  This numerical method can be
easily reconstructed to study another strong correlated system in
the future research.

\begin{figure}[t]
 \centering
\includegraphics[width=7cm]{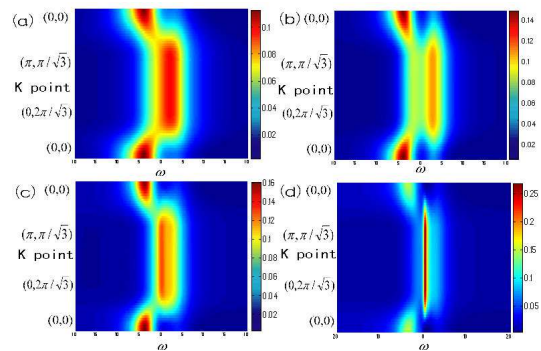}\hspace{0.5cm}
\caption{\label{fig:epsart}(Color online) The K-dependent spectral
function $A_k(\omega)$ for different temperature, when $U/t=7.0$.
$t$ is the kinetic energy in equation (2). (a) At $T/t=1.67$, the
red part shows a Fermi-liquid-like peak near Fermi energy. (b) At
$T/t=1.11$, a narrow pesudogap shown by green region. Two peak
found around the pesudogap is shown by yellow and orange color.
(c) At $T/t=0.5$, a Fermi-liquid-like peak appears again shown by
red color. (d) At $T/t=0.2$, an obvious Kondo peak appears shown
by red color.}
\end{figure}

\section{Phase Diagram}
We investigate the double occupancy
$D_{occ}=\partial{F}/\partial{U}=\frac{1}{4}\sum_{i}\langle{n_{i\uparrow}n_{i\downarrow}}\rangle$
as a function of interaction $U$ for various temperature, where F
is the free energy. When the interaction is lower than $U/t=8.6$,
the $D_{occ}$ increases as the temperature decreases due to the
enhancing of the itinerancy of atoms, as shown in Fig. 3. When the
interaction is stronger than $U/t=8.6$, the effect of the
temperature on $D_{occ}$ is weakened. The $D_{occ}$ decreases as
the interaction increases due to the suppressing of the itinerancy
of the atoms. When the interaction is stronger than the critical
interaction of the Mott transition, $D_{occ}$ for different
temperature is coincident, which shows the temperature does not
affect the double occupancy distinctly. The continuity of the
evolution of the double occupancy by interaction shows that it is
a second order transition.

We employ the maximum entropy method \cite{Gubernatis} to
calculate the density of states (DOS) which describes the number
of states at frequency $\omega$. Fig. 4(a) shows DOS for different
interaction at $T/t=0.5$. There is a Fermi-liquid-like peak when
$U/t=4.0$. A pseudogap formed by the splitting of
Fermi-liquid-like peak appears when the interaction increases,
such as $U/t=8.0$ and $U/t=9.0$. When the interaction is stronger
than the critical interaction $U_c/t=11.6$, such as $U/t=12.0$ and
$U/t=20.0$, the system becomes an insulator indicated by an opened
gap. DOS for different interaction at $T/t=0.25$ is shown in Fig.
4(b). When $U/t=4.0$, there is also a Fermi-liquid-like peak. When
the interaction increases, such as $U/t=8.0$, a Kondo resonance
peak appears which is shown by a sharp quasi-particle peak with
two shoulders. When $U/t=9.3$, the Kondo resonance peak is
suppressed and a pseudogap appears, which is the intermediary
state between the Fermi liquid and the Mott insulator. A gap is
opened when the interaction is stronger than the critical
interaction $U_c/t=10.6$. Instead of directly splitting into two
parts shown in Fig. 4(a), a Kondo peak appears, which is formed by
the effect between the local atom and the itinerant atom at low
temperature, as shown in Fig. 4(b).

\begin{figure}[t]
\centering
\includegraphics[width=8cm]{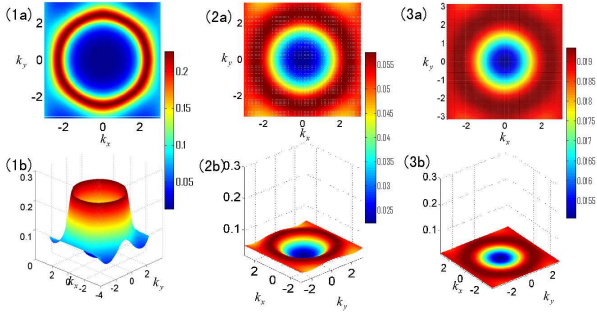}\hspace{0.5cm}
\caption{\label{fig:epsart}(Color online) The Fermi surface as a
function of momentum $k$ for different interaction at $T/t=1.25$:
(1) $U/t=5.0$, (2) $U/t=10.0$, (3) $U/t=16.0$. $t$ is the kinetic
energy in equation (2).}
\end{figure}

We could also get the K-dependent spectral function
$A_k(\omega)=-ImG_k(\omega+i0)/\pi$, which describes the
distribution probability of the quasi-particle with momentum $k$
and energy $\omega$. Fig. 5 shows $A_k(\omega)$ for different
temperature, where $U/t=7.0$. At $T/t=1.67$, there exists a
quasi-particle peak which shows a metallic behavior, as shown in
Fig. 5(a). In Fig. 5(b), a pseudogap appears at $T/t=1.11$, which
is formed by the splitting of the quasi-particle peak. When
$T/t=0.50$, the pseudogap disappears and there is a quasi-particle
peak again, as shown in Fig. 5(c). An obvious Kondo peak appears
when $T/t=0.2$, as shown in Fig. 5(d). It shows a reentrant
behavior in the transition between Fermi liquid and pseudogap when
the interaction is fixed and the temperature decreases. This
behavior is based on the Kondo effect which suppresses the
splitting of the quasi-particle peak at low temperature.

We study the Fermi surface as a function of momentum $k$ by
$A(k;\omega=0)\approx-\frac{1}{\pi}\lim_{w_n\rightarrow0}Im
G(k,i\omega_n)$. A linear extrapolation of the first two Matsubara
frequencies is used to estimate the self-energy to zero frequency
\cite{Parcollet}. Fig. 6 shows the Fermi surface for different
interaction at $T/t=1.25$. A circular ring which means the
particles distribute on a certain energy displays a metallic
behavior, as shown in Fig. 6(1a). As the interaction increases,
the ring becomes bigger, as shown in Fig. 6(2a). When the
interaction is stronger than the critical interaction $U/t=13.7$,
the Fermi surface becomes a nearly flat plane, as shown in Fig.
6(3a). From Fig. 6(1b) to Fig. 6(3b), we can find the amplitude of
the spectral weight becomes small and the breadth become wider.
The Fermi surface translates from a determined surface into a flat
plane due to the localization of the particles when the
interaction increases.

\begin{figure}[t]
\centering
\includegraphics[width=7cm]{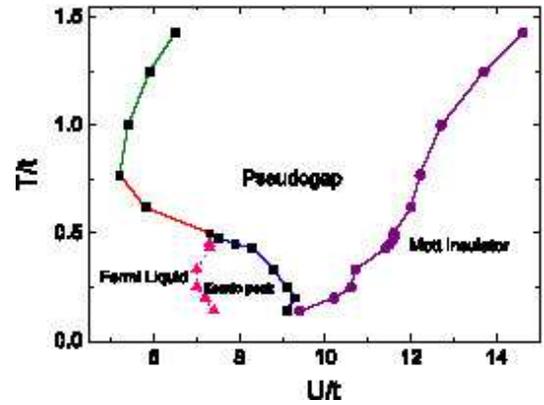}\hspace{0.5cm}
\caption{\label{fig:epsart}(Color online) The phase diagram of
Fermi atoms in triangular optical lattice, where the square plots
with solid line (green, red, blue) indicate the transition line of
the Fermi liquid and the pseudogap, the circular plots with solid
line (purple) indicate the Mott transition line, the triangular
plots with dash line (pink) mark the Kondo peak appearing region.
$t$ is the kinetic energy in equation (2). }
\end{figure}

The phase diagram of cold atoms trapped in triangular optical
lattice is shown in Fig. 7, where $a_s$ indicates the s-wave
scattering length. The translation between the Fermi liquid and
the pseudogap shows a reentrant behavior. For a fixed interaction
weaker than $U/t=7.2$, when the temperature decreases, the system
translates from Fermi liquid to pseudogap. When the temperature is
lower than the critical temperature distributing on the red solid
line, the system translates from pesudogap to Fermi liquid. There
is a Kondo peak region, which is signed by the blue solid line and
the pink dash line. If temperature is lower than $T/t=2.0$, a
Kondo peak emerges before the appearing of the pseudogap when the
interaction increases. When the interaction is stronger than the
critical interaction of Mott transition distributing on the purple
line, the system translates from pseudogap to insulator confirmed
by an opened gap.

\section{Experimental Protocol}
We design an experiment to investigate the quantum phase
transition in triangular optical lattice. The experimental
protocol can be taken as follows: The $^{40}K$ atoms can be
firstly produced to be a pure fermion condensate by evaporative
cooling \cite{Ohara}. Three laser beams at wavelength
$\lambda=1064$ nm are used to form the triangular optical lattice
\cite{Tung} to trap $^{40}K$ atoms. The lattice depth $V_0$ is
used to adjust the kinetic energy $t$ and the interaction $U$. The
on-site interaction can be adjusted by Feshbach resonance
\cite{Loftus,Martin, Regal, Klempt, Thilo}. The s-wave scattering
length is used to determine the effective interaction. The
temperature can be extracted from time-of-flight images by means
of Fermi fits in experiment \cite{Schneider}.

When the interaction increases, the system translates from Fermi
liquid ($a_s<48a_0$) to Mott insulator ($a_s>111a_0$) at
temperature $T=5.96nK$, when $V_0=10E_r$. In order to detect the
double occupancy $D_{occ}$, we increase the depth of optical
lattice rapidly to prevent the further tunnelling first. Then, we
shift the energy of atoms on doubly occupied sites by approaching
a Feshbach resonance. And then, one spin component of atoms on the
double occupied sites is transferred to a unpopulated magnetic
sublevel by using a radio-frequency pulse. The double occupancy
can be deduced by the fraction of transformed atoms obtained by
the absorption imagining \cite{Jordans, Strohmaier}. At
$T=5.96nK$, $D_{occ}$ decreases from $0.11009$ ($a_s=45a_0$) to
$0.00456 $ ($a_s=130a_0$) with increasing atomic interaction.

By ramping down the optical lattice slowly enough, the atoms stay
adiabatically in the lowest band while the quasi-momentum is
approximately conserved. Then, the optical lattice is converted
from a deep one into a shallow one and the quasi-momentum is
preserved. After completely turning off the confining potential,
the atoms ballistically expand for several milliseconds. Then by
the absorption imagining, one can get the Fermi surface
\cite{Michael, Chin}. At $T=5.96nK$, the circular ring shape of
the Fermi surface when $a_s=41a_0$ transforms into a flat plane
when $a_s=162a_0$.

\section{Summary}
In summary, we investigate the Mott transition of the cold atoms
in 2D triangular optical lattice set up by three laser beams. The
system evolves from Fermi liquid into Mott insulator for
increasing interaction, and a reentrant behavior of the transition
between Fermi liquid and pesudogap is found, due to the Kondo
effect. Our study presents a helpful step for understanding the
strong correlated effect in the frustrated system, such as the
spin liquid. Beyond DMFT, DCA is improved to incorporated the
nonlocal correlation which can not be simply ignored in the
frustrated system. This numerical method is universally to
investigate strongly correlated system, such as the
high-temperature superconductor and the colossal
magnetoresistance.

This work was supported by NSFC under grants Nos. 10874235,
10934010, 60978019, the NKBRSFC under grants Nos. 2006CB921400,
2009CB930701 and 2010CB922904.

\end{document}